\documentclass[prl,twocolumn,preprintnumbers,amsmath,amssymb,nofootinbib,epsfig,bmamsfonts,yfonts,superscriptaddress]{revtex4}
\usepackage{graphicx}
\usepackage{float}
\usepackage{amsmath}
\usepackage{color}
\relax

\begin{document}

\preprint{CERN-TH-2019-120}
\title{What attracts to attractors?}

\author{Aleksi Kurkela}
\email{a.k@cern.ch}
\affiliation{Theoretical Physics Department, CERN, CH-1211 Gen\`eve 23, Switzerland}
\affiliation{Faculty of Science and Technology, University of Stavanger, 4036 Stavanger, Norway}
\author{Wilke van der Schee}
\email{wilke.van.der.schee@cern.ch}
\affiliation{Theoretical Physics Department, CERN, CH-1211 Gen\`eve 23, Switzerland}
\author{Urs Achim Wiedemann}
\email{urs.wiedemann@cern.ch}
\affiliation{Theoretical Physics Department, CERN, CH-1211 Gen\`eve 23, Switzerland}
\author{Bin Wu}
\email{b.wu@cern.ch}
\affiliation{Theoretical Physics Department, CERN, CH-1211 Gen\`eve 23, Switzerland}

%\preprint{CERN-TH-2019-120}

\begin{abstract} 

Whether, how, and to what extent solutions of Bjorken-expanding systems become insensitive to aspects of their initial conditions is of importance for heavy-ion collisions. Here we study 1+1D and phenomenologically relevant boost-invariant 3+1D systems in which initial conditions approach a universal attractor solution. 
In Israel-Stewart theory (IS) and kinetic theory where the universal attractor extends to arbitrarily early times, we show that all initial conditions approach the attractor at early times  
by a power-law while their approach is exponential at late times.
In these theories, the physical mechanisms of hydrodynamization operational at late times do not drive the approach to the attractor at early times, and the early-time attractor is reached prior to hydrodynamization.
 In marked contrast, the attractor in strongly coupled systems is realized concurrent with hydrodynamization.
This qualitative difference may offer a basis for discriminating weakly and strongly coupled scenarios of heavy-ion collisions. 

\end{abstract}

\maketitle
\begin{figure}[ht]
%\begin{center}
 \includegraphics[width=0.45\textwidth]{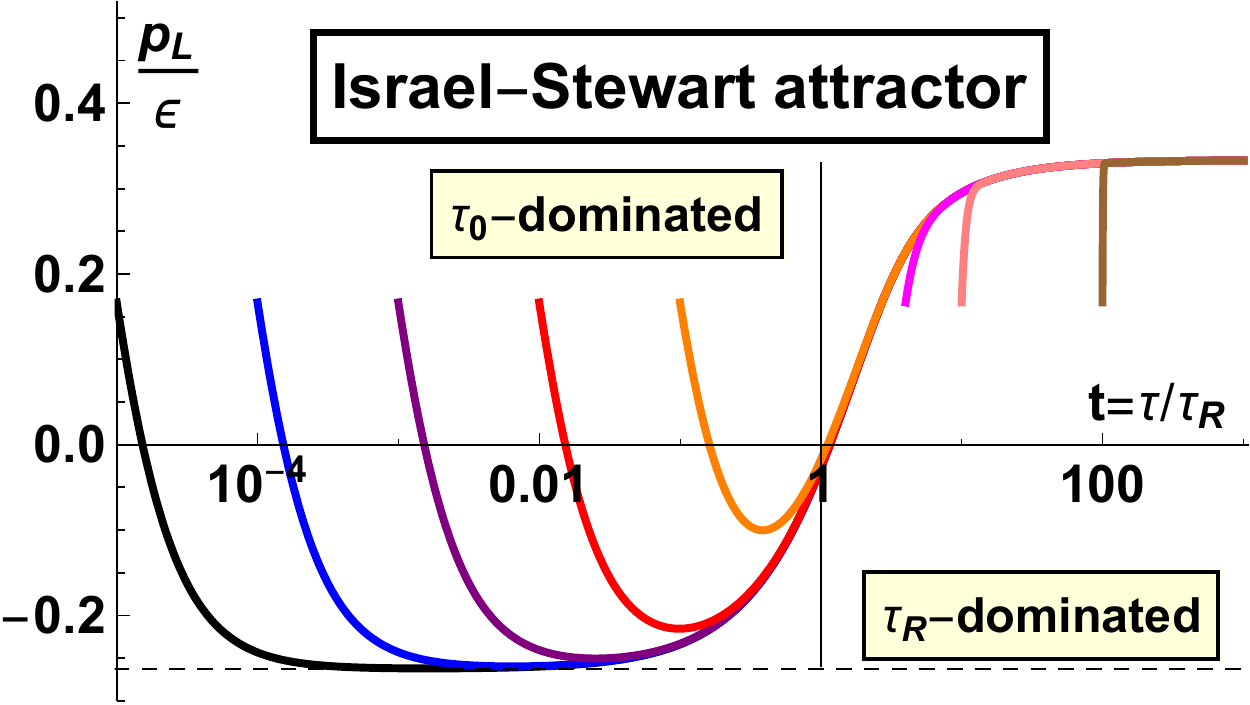}
 \hfill
  \includegraphics[width=0.45\textwidth]{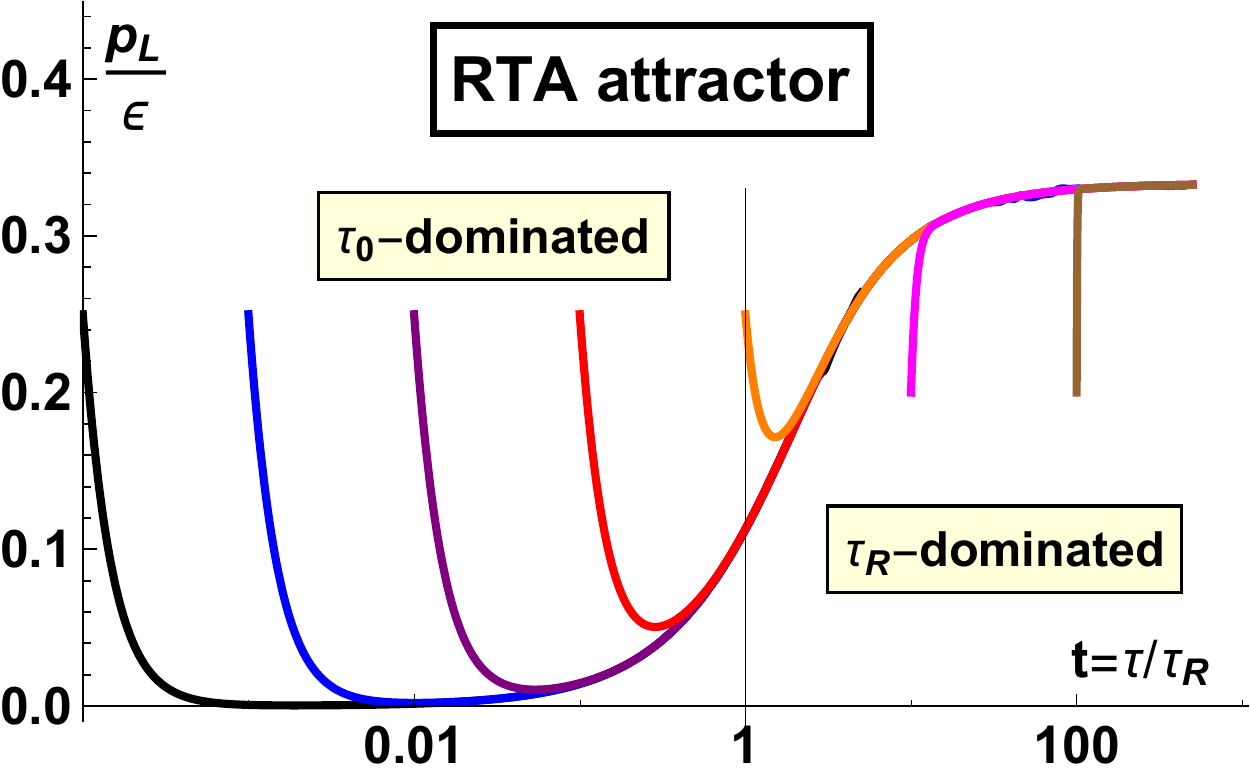}
 \hfill
   \includegraphics[width=0.45\textwidth]{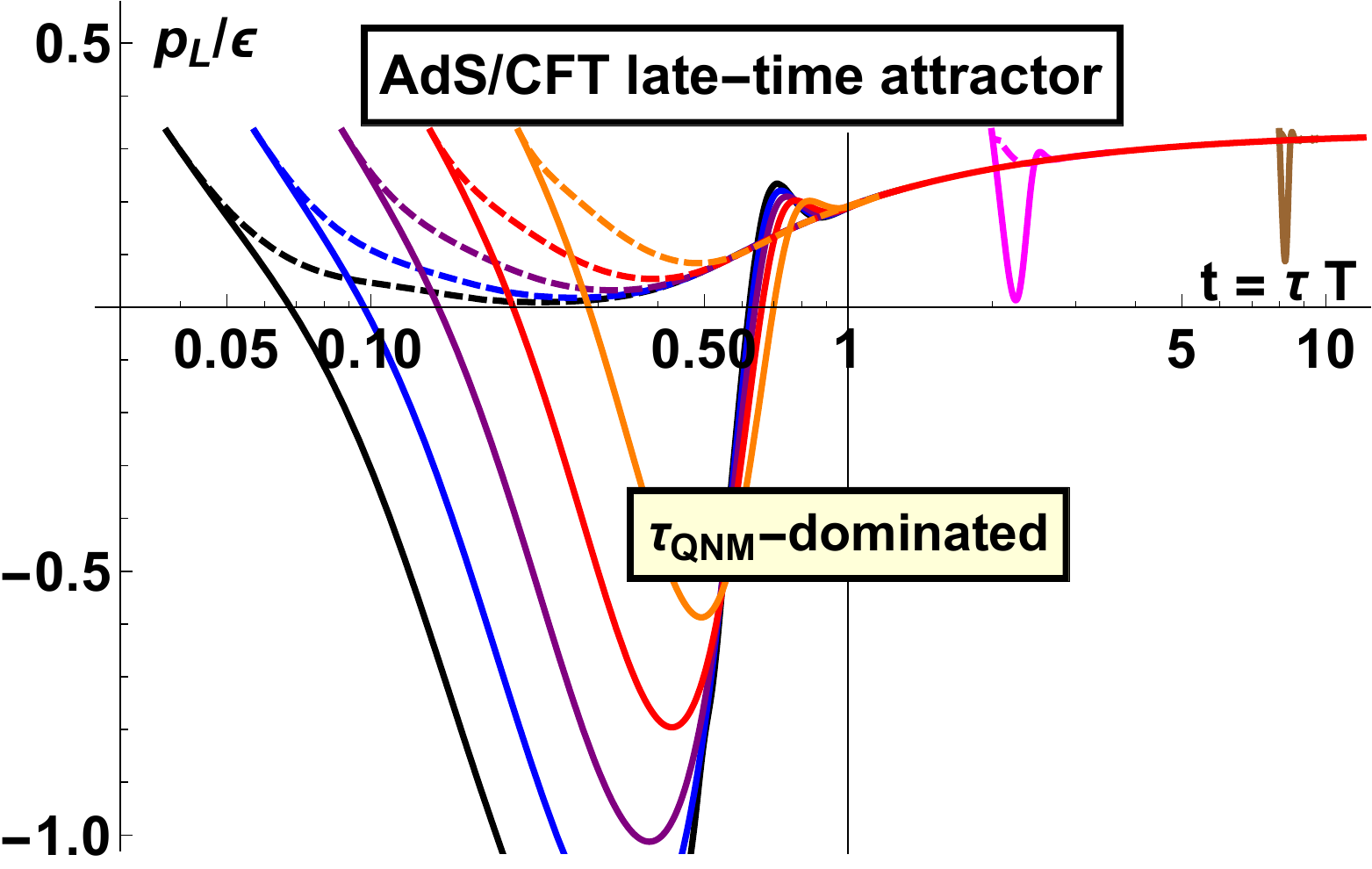}
%\end{center}
  \caption{Approach to the attractor in different theories.  
  Different lines correspond to out-of-attractor initializations at different times $t_0$. 
  Upper and middle panel: For $t < 1$, the slope of the
  approach remains constant on a log-linear scale, indicating that the time-dependence is set by the initialization time $t_0$ and thus governed by the expansion rate. 
  In contrast, for $t>1$, the approach to the attractor appears on a log-linear scale  steeper and steeper with increasing $t$,
  indicating its dependence on the interaction rate $\tau_R$ which does not depend on $t_0$. 
  Lower panel: Qualitatively different behavior is seen for $\mathcal{N}=4$ SYM theory, where information characteristic of specific 
  initial condition is lost only at times $t \geq 1$, irrespective of how early the system is initialized. }
\label{fig1}
\end{figure}

In a dynamical system,  an attractor is the particular solution to which arbitrary initial conditions within the basin of attraction relax at sufficiently late times. 
 In general, the attractor is characterized by the competition between the expansion rate that drives the system towards local anisotropy, and the isotropizing interaction rate~\cite{Kurkela:2011ub}.  
Attractors are easily found empirically by evolving a set of different initial conditions (see Fig.~\ref{fig1} for an example). 
Recently, such attractor solutions have received attention in the context of ultra-relativistic heavy-ion collisions.
Their form is of interest for understanding the onset of fluid-dynamic behavior~\cite{Janik:2006gp,Beuf:2009cx,Heller:2011ju,Heller:2013fn,Heller:2015dha,Keegan:2015avk,Denicol:2016bjh,Heller:2016rtz,Romatschke:2017vte,Spalinski:2017mel,Strickland:2017kux,Behtash:2017wqg,Romatschke:2017acs,Blaizot:2017ucy,Heller:2018qvh,Denicol:2018pak,Spalinski:2018mqg,Behtash:2019txb,Strickland:2019hff,Kurkela:2015qoa,Strickland:2019jut}
and  the origin of the non-thermal fixed-point behavior in far-from-equilibrium dynamics~\cite{Kurkela:2011ub,Kurkela:2015qoa,Berges:2013fga,Berges:2014bba,Mazeliauskas:2018yef,Boguslavski:2019fsb}. 
For the phenomenology of heavy-ion collisions, these studies are needed to clarify to what extent different observables inform us either about the details of the initial conditions or about the material properties of the system.

Whether an attractor solution exists at arbitrarily early times depends on the dynamics that drives the initial conditions to the attractor. 
Here we point out that some models undergoing Bjorken expansion do exhibit attractor behavior at 
arbitrarily early times while others don't. The existence of the early-time attractor is a consequence of the longitudinal expansion at early times which would 
render heavy-ion phenomenology insensitive to the unknown details of the longitudinal structure of the initial state.

%%%%%%%%%%%%%%%%%%%%%%%%%%%%%%%%%%%%%%

\underline{Israel-Stewart theory:} In Bjorken-expanding Israel-Stewart (IS)  theory \cite{Israel:1979wp} with transverse translational symmetry, an attractor solution exists for the ratio of longitudinal pressure $p_L$ over energy density $\varepsilon$,
\begin{eqnarray}
	\partial_{\tau} \varepsilon
        +\frac{1}{\tau}  \left[\varepsilon+p_L \right] &=& 0\, , \label{eq1}\\
%        \end{equation}
%        \begin{equation}
        \partial_\tau \phi+\frac{4}{3}\frac{\phi}{\tau}&=&-\frac{1}{\tau_{\rm R}}\left[\phi-\frac{4}{3}\frac{\eta}{\tau} \right]\, .
        \label{eq2}
\end{eqnarray}
Here $\tau$ is the proper time and $\phi \equiv \frac{1}{3}\varepsilon-p_L$. The time governing the relaxation to fluid-dynamic constitutive equations is $\tau_{\rm R} = \textstyle\frac{5}{a}
 \textstyle\frac{\eta}{(\varepsilon+P)\, }$, where $a$ is a free parameter, conventionally fixed to $a=1$. This is the timescale on which linearized non-hydrodynamic excitations decay.

We work for a conformal equation of state $\varepsilon = 3P$
and constant specific viscosity $\eta/s \propto \eta/\varepsilon^{3/4}$. The equation of motion for the ratio $x \equiv p_L/\varepsilon$ 
(written for convenience  in the rescaled time $t = \tau/\tau_{\rm R}$)
reads 
\begin{equation} 
 \left(\frac{3}{4}-\frac{x}{4}\right) \frac{d x}{d t}=\frac{45 x^2-30 x+ 5+ \left( 15 t  (1-3 x)-16a\right)}{45 t }\, .
 \label{eq3}
 \end{equation}

The limit $a \to 0$ at finite $\tau_{\rm R}$ is equivalent to an ideal IS theory with $\eta = 0$. In this simplest case, the attractor is the equilibrium $x_{\rm A} = 1/3$, and how the attractor is
approached is given by rewriting (\ref{eq3}) in terms of the deviation $\delta = x-x_{\rm A}$
 \begin{equation}
\left( \frac{3}{4}- \frac{ 1+ 3 \delta}{12} \right)\frac{d \delta}{d t}=
\frac{ \delta^2}{t}- \delta\, .
\label{eq4}
\end{equation}
Depending on whether $t > \delta$ or $t< \delta$, the approach to the attractor is governed by the expansion rate ($t^{-1}$) or by
the interaction rate (independent of $t$), respectively. 
At all times,  sufficiently small deviations from the attractor decay exponentially $\delta \sim e^{-3 t /2 }$, which is characteristic for 
linearized non-hydrodynamic perturbations around thermal 
equilibrium~\cite{Heller:2015dha}. The factor $3/2$ arises from the non-trivial time-evolution of the background. 

\begin{figure*}[t!]
\begin{center}
 \includegraphics[width=0.31\textwidth]{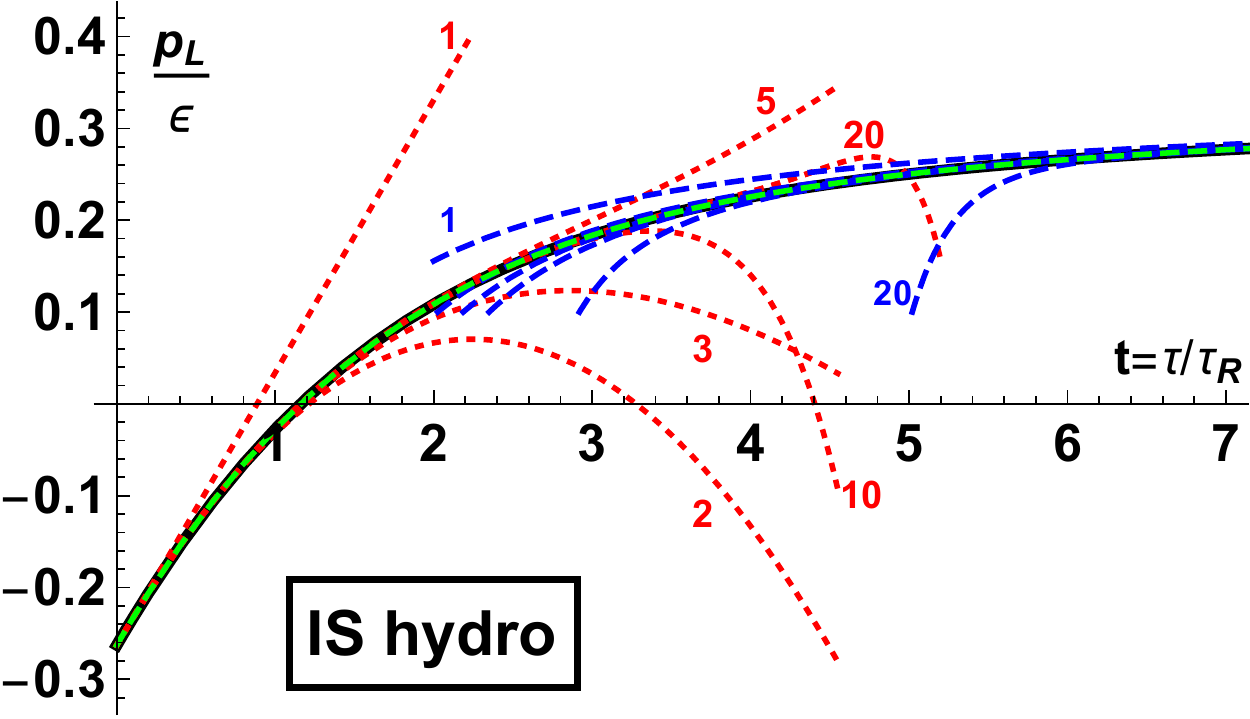}
 \hfill
 \includegraphics[width=0.31\textwidth]{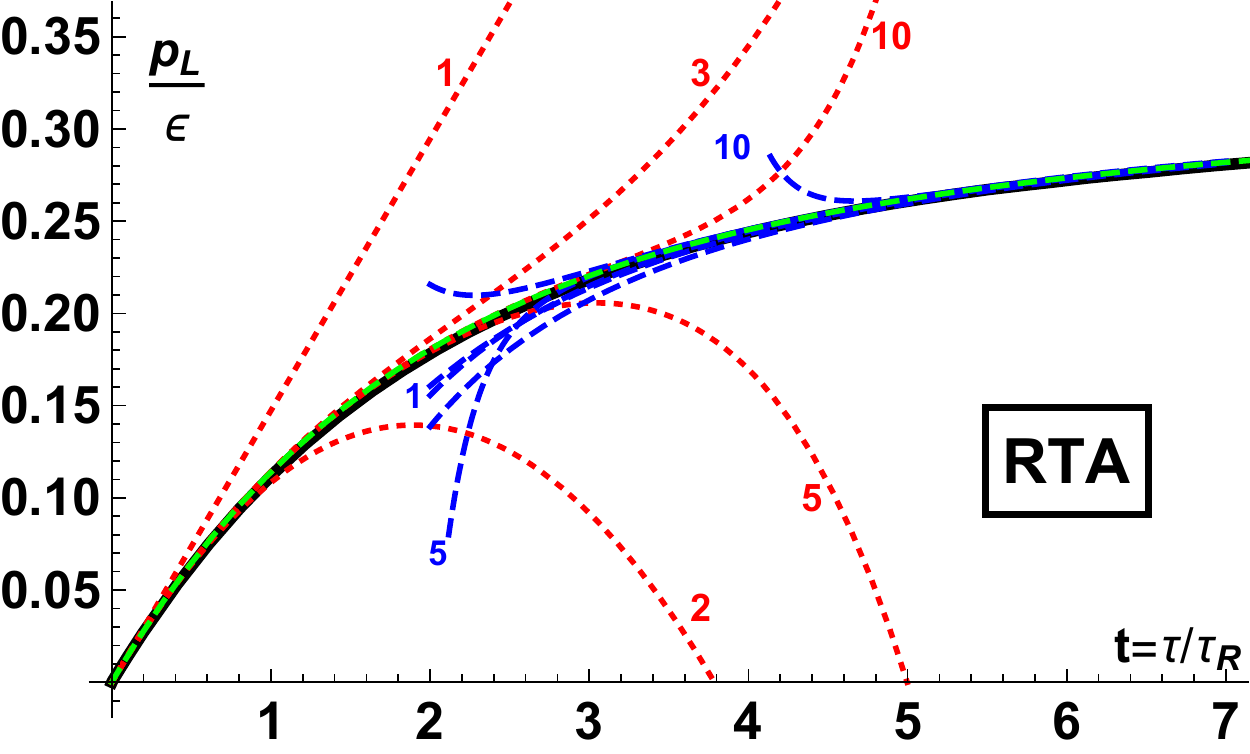}
 \hfill
   \includegraphics[width=0.31\textwidth]{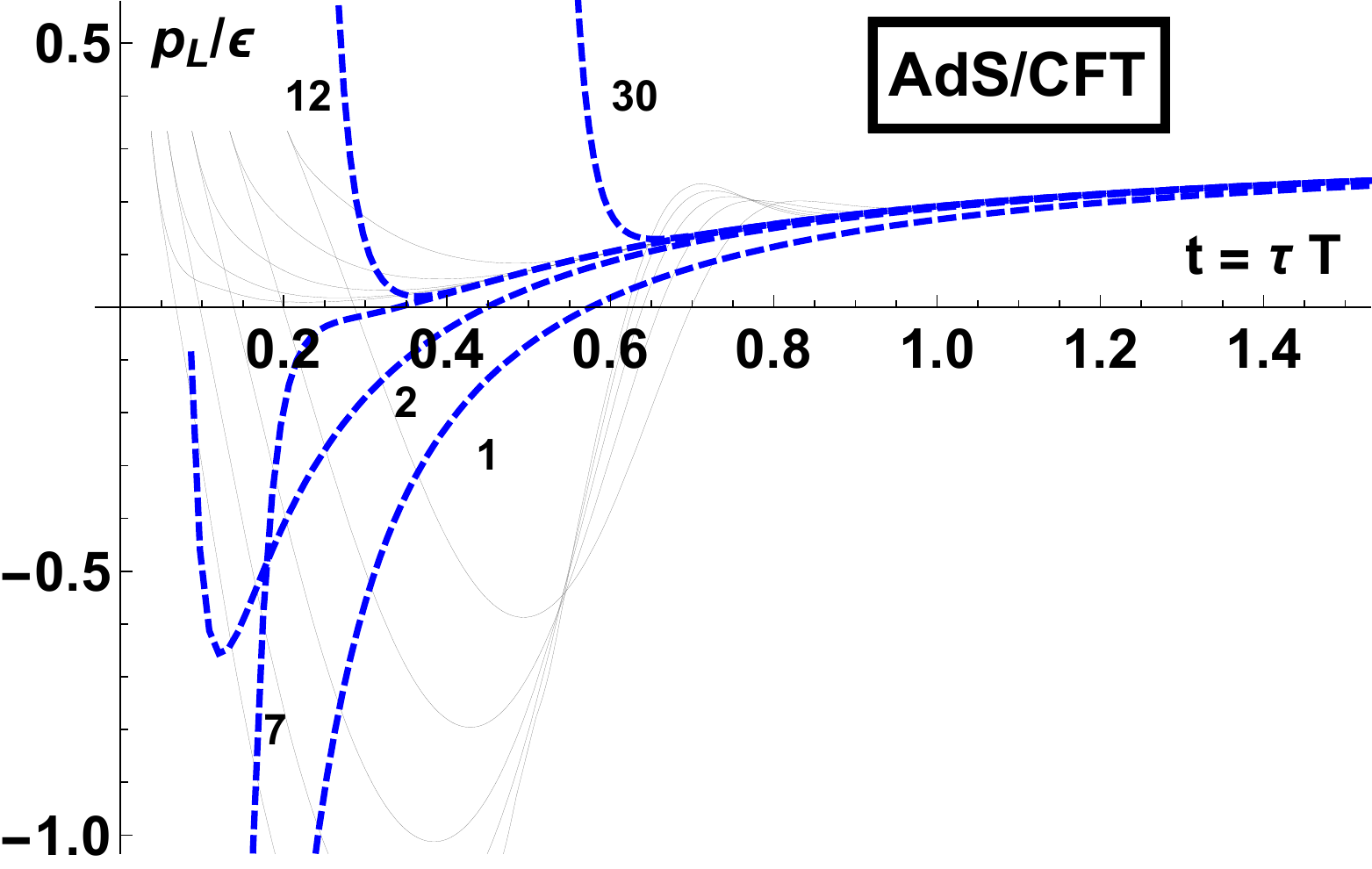}
\end{center}
\vspace{-.5cm}
  \caption{Early-time (red-dashed) and late-time (blue-dashed) expansions of the attractor solutions (black). 
  % of 1+1D Israel-Stewart theory (left panel), kinetic theory (middle panel) and ${\mathcal N}=4$ SYM (right panel). 
  Orders of the expansions are given by numbers in the plots. 
The Pad\'e approximant (green-dashed) extends the early-time expansion to any finite $t$ in systems where the early-time attractor exists.
For ${\mathcal N}=4$ SYM, grey curves are the specific solutions shown in Fig.~\ref{fig1}.}
\label{fig2}
\end{figure*}

For finite $a$, eq.~(\ref{eq3}) corresponds to the first-order IS theory, which 
has two solutions that remain regular for $t_0 \to 0$ with limits  $\lim_{t_0 \to 0} x_{\pm} (t_0) = \frac{1}{15} \left(5 \pm 4 \sqrt{5a} \right)$, respectively. 
The solution $ x_{-} (t)$ is the attractor solution $x_{\rm A}(t)$, while $x_{+} (t)$ limits the basin of attraction from above%
\footnote{
We note that within IS theory, $a$ is a free parameter. The choice 
$a=1$ amounts to equating $\tau_R$ to the second order hydrodynamic
coefficient $\tau_\pi$ of RTA kinetic theory. The choice $a=5/16$ would instead ensure that the early-time
attractor of IS theory coincides with that of kinetic theory, $x_-(t_0)=0$.
}. 
While we do not have an analytic solution $x_-(t)$, the attractor can be expanded 
at late and early times, see Supplemential Material for details. The main finding, see Fig.~\ref{fig2}, is that the early-time expansion is a convergent series that can be analytically continued to arbitrarily late times using standard techniques, while the late-time expansion is a non-convergent,
asymptotic Borel-resummable series.
We next discuss the transient dynamics that evolves generic initial conditions from time $t_0$  towards the attractor. At late initializations, $t_0 \gg 1$, this is the well-known exponential decay of linearized
non-hydrodynamic modes governed by (\ref{eq4}) that, as discussed above, is determined by the interaction rate, see Fig.~\ref{fig1}.  Exponential decays with this timescale have been revealed in 
Borel-resummations of the late time  expansion~\cite{Heller:2015dha}. 
In marked contrast, at early times, (\ref{eq3}) becomes  
 \begin{equation}
\left(\frac{3}{4}-\frac{x_{\rm A}(t=0) + \delta}{4}\right) \frac{d \delta}{d t}=  
 \frac{\delta}{t} \left[ \frac{ 15\, \delta -8 \sqrt{5 a}}{15 }\right]\, ,
 \end{equation}
and 
the attractor is approached by a power law 
\begin{equation}
  \delta \sim t^{-\frac{8 \sqrt{5a }}{\sqrt{5a}+10}}\, .
 \end{equation}
 We emphasize that the timescale of this decay becomes increasingly rapid and ultimately instantaneous with decreasing $t_0$ (see Fig.~\ref{fig1}). 
 This is the hallmark of a decay governed by the expansion rate. It is qualitatively different from what one expects from the decay of non-hydrodynamic modes, 
 and it forces the decay to the attractor prior to hydrodynamization. 

Higher-order fluid-dynamic models like rBRSSS~\cite{Baier:2007ix} amount to replacing in (\ref{eq2}) the relaxation to the first order constitutive
relation (the term $\textstyle\frac{4}{3}\textstyle\frac{\eta}{s}$) by relaxation to the second order one. In general, all additional terms thus introduced are $\propto \textstyle\frac{1}{\tau}$.  
As a consequence, the value of the early-time attractor changes, but the early-time power-law approach of arbitrary initial conditions towards the attractor is unaffected.

\underline{Kinetic theory}: Features similar to the above can also be seen in Bjorken-expanding massless kinetic theory in the relaxation time approximation (RTA)
\begin{equation}
\partial_{\tau} f+\vec{v}_{\perp} \cdot \partial_{\vec{x}_{\perp}} f-\frac{p_{z}}{\tau} \partial_{p_{z}} f=- \frac{\left(-v_{\mu} u^{\mu}\right)}{\tau_{\rm R}}\left[ f - f_{\rm eq}\right]\, .
\label{eq10}
\end{equation}
Here, the distribution function \(f\left(\tau, \vec{x}_{\perp} ; \vec{p}_{\perp}, p_{z}\right)\) relaxes to equilibrium $f_{\rm eq}$. It depends on
 \(p^{\mu}=\left(p, \vec{p}_{\perp}, p_{z}\right), p= \sqrt{ \vec{p}_{\perp}^{2}+p_{z}^{2}},\) and on the proper time
\(\tau\); $u_\mu$ denotes the rest frame of the energy density and
 \( \vec{v}_{\perp}=\vec{p}_{\perp} / p, v_{z}=p_{z} / p\) are transverse and
longitudinal velocities, respectively. We work with a conformal relaxation time $\tau_{\rm R}^{-1}= \gamma \varepsilon^{1/4}$.

For systems with transverse translational symmetry, this Boltzmann equation can be reduced to a tower of moment equations~\cite{Denicol:2016bjh,Strickland:2019hff} describing the time evolution of various integral moments of the distribution function $p_{l} \equiv \int_{-1}^{1} \frac{d v_{z}}{2} \int \frac{4 \pi d p p^{3}}{(2 \pi)^{3}} f  v_{z}^{2l} $. Energy density and longitudinal pressure correspond to the first two moments,  $\varepsilon=p_0 $, $p_L = p_1$.  The first two equations in the hierarchy result in 
\begin{equation}
	\left(\frac{3}{4}-\frac{x}{4}\right) \frac{d x}{d t}=
	\frac{3 x^2-6 x+3 y+t(1-3x)}{3 t}	\, ,
\label{eq12}
\end{equation}
where $y \equiv p_2/\varepsilon$; see Supplemental Material for further details. The attractor is found amongst the regular solutions.
Solutions that remain regular for $t_0\to 0$ satisfy   $\lim_{t_0 \to 0} x_{\pm} (t_0) = 1 \pm \sqrt{1-y(0)} $. All physical values 
$x\leq 1$ lie within the basin of attraction since $x_+ > 1$. Since $0< y < x$, the attractor solution at early times is   $x_{\rm A} (0) = 0$, and therefore also all higher moments
$p_l(0)$ vanish. At late times, it follows trivially
from (\ref{eq12}) that the attractor approaches equilibrium, $\lim_{t \to \infty} x_{\rm A} (t) = \textstyle\frac{1}{3}$.

The late-time fluid-dynamic expansion of this transport theory has been computed to high orders~\cite{Heller:2016rtz}. Similar to IS theory, 
it is a non-convergent, asymptotic Borel-resummable series. Also, similarily, the early-time expansion 
has a finite radius of convergence and can be extended to arbitrary late times by standard techniques (see Supplemental Material). 

The transient dynamics according to which generic initial conditions approach this attractor shares the main qualitative features of the IS theory discussed above, see Fig.~\ref{fig1}. 
At late initializations, $t_0 \gg 1$, eqs.(\ref{eq12}) and (\ref{eq4}) govern identical exponential decays of linearized non-hydrodynamic modes.  For early times, $t_0 \ll 1$, the
decay of $\delta = x-x_A$ to the attractor depends on $y$, and through $y$ on the initial conditions of all higher moments. Because $6x> 3y$ for any system, an approximate
solution of the approach to the attractor can be obtained for a generic initial condition by neglecting $y$ in (\ref{eq12}) which leads to the power law decay 
$\delta \sim t^{-8/3}$. Similar reasoning suggests that $y$ would approach its attractor  $\sim t^{-16 /3}$ thus justifying the above approximation; in fact, the same reasoning gives for all higher orders 
$p_l(t)/\varepsilon(t) \propto x(t)^l$. These power laws are easily seen in numerical solutions of eq.~(\ref{eq10})  (see Supplemental Material).

\underline{Strongly coupled $\mathcal{N}=4$ SYM:} The third class of qualitatively different models of collectivity invoked in heavy-ion physics is given by strongly coupled quantum field theories with known gravity duals. Here, we contrast and compare the early-time dynamics 
in strongly coupled $\mathcal{N}=4$ Super Yang-Mills (SYM) theory with the attractor behavior observed above. 
% Here, we ask to what extent in strongly coupled Super Yang-Mills (SYM) theory, the approach to an attractor exhibits the same transition  from an expansion-driven early-time to an interaction-driven late-time dynamics. 
The time evolution of $p_L/\epsilon$ can be solved for boost invariant initial conditions using standard methods of holography, \emph{i.e.}, by solving the 5-dimensional Einstein equations with the ansatz for the line-element
\begin{equation}
ds^2 = - 2 \rho^{-2} d\rho d\tau - A d\tau^2 +S^2 e^B d{\bf x}_\perp^2 + S^2 e^{- 2 B}d\xi^2\, ,
\end{equation}
with $\xi$ the space-time rapidity and $\rho$ the internal fifth dimension. The initial conditions are specified by the 
$\rho$-dependent function $B(\tau_0,\rho) = B_{\rm AdS}(\tau_0,\rho) + B_0(\rho)$, with $ B_{\rm AdS}(\tau_0,\rho) =-\frac{2}{3}\log(\tau_0+\rho)$ the vacuum AdS solution; $B(\tau, \rho)$, $A(\tau, \rho)$ and $S(\tau, \rho)$ then follow from the Einstein equations \cite{Chesler:2009cy}. Unlike in the models above\footnote{We recall that also 
kinetic theory  can be initialized with different classes of initial conditions by varying the values of higher moments at $t_0$, but all initial conditions approach the same attractor on
a timescale $t_0$,
see Supplemental Material. In IS theory, there are no further degrees of freedom beyond 
$\textstyle\frac{p_L(t_0)}{\varepsilon(t_0)}$ that can be specified.} changing the initial Cauchy data
$B_0(\rho)$ amounts to a choice not only for
$\textstyle\frac{p_L(t_0)}{\varepsilon(t_0)}$, but also for all its derivatives at $t_0$~\cite{Beuf:2009cx, Heller:2012je}. 
We therefore study two different families of initial conditions, 
 $B_0^{(UV)}(\rho) = e^{-40 \rho T} 32 \rho^5 T^5$ and $B_0^{(IR)}(\rho) = 32 \rho^5 T^5$, where $T$ is the effective temperature determined from the energy density.
While their functional form is somewhat arbitrary, %and partially motivated to stay on the same plot range, 
they are chosen such that their initial anisotropy $\textstyle\frac{p_L(t_0)}{\varepsilon(t_0)}$ and its first derivative are equal. 
They differ qualitatively in that their support  is either localized  close to the boundary (UV) or spread out in the fifth dimension (IR). 

Fig.~\ref{fig1} shows solutions in which both initial conditions are evolved from a set of different initialization times $t_0$. 
In marked contrast to IS theory and kinetic theory, different initial conditions do not reach a unique curve on time scale $t_0$. Rather, 
information about the initial condition is lost only at times $t\sim 1$, and only on that time scale 
solutions converge to a common attractor. By closer inspection of these results (data not shown) we observe that
both for early initializations $t_0 < 1$ and for late initializations $t_0 > 1$, differences between solutions show the 
oscillatory behavior characteristic for the decay of quasi-normal modes (QNM)~\cite{Spalinski:2018mqg} that are exponentially damped with 
time scale $\tau_{\rm QNM}$. These solutions reach a unique attractor only at late times $t > 1$. In this sense, only the late time attractor is 
universal in $\mathcal{N}=4$ SYM; this is consistent with Fig.2 of Ref.~\cite{Romatschke:2017vte}.

It is curious to note that curves initialized with $B^{(UV)}_0(\rho)$ approach the high-order hydrodynamic late-time expansion \cite{Heller:2013fn} significantly earlier 
than curves initialized with $B^{(IR)}_0(\rho)$, see Fig.~\ref{fig2}. The latter initial condition is expected to give rise to a larger connected  two-point function $\langle T^{\alpha \beta} T^{\mu \nu}\rangle$ than the 
former~\cite{Ecker:2015kna}. We believe that this observation, together with vanishing n-point functions in the above kinetic theory and in IS theory, should motivate further research into the relation of higher connected n-point functions and attractor behavior. This question could be asked not only in   $\mathcal{N}=4$ SYM, but also in BBGKY-extensions of the Boltzmann equation.

\underline{Attractors in boost-invariant 3+1D kinetic theory:} 
Would an early-time attractor, if it exists, leave observable imprints? If so, this could provide a tool for disentangling qualitatively different 
microscopic candidate theories of weakly or strongly coupled quark gluon plasma. With this motivation, we now ask which
aspects of the attractor behavior are accessible in collisions with a finite transverse extent and realistic transverse gradients. 
%The early time attractor of IS theory reflects ad hoc assumptions, and the early-time attractor of  ${\cal N}=4$ SYM does not exist. 
We focus on the kinetic theory (\ref{eq10}) as it possesses an early-time attractor. We have solved (\ref{eq10})
for realistic initial transverse profiles~\cite{Kurkela:2019kip}. 
Because the early-time approach to the attractor is governed by the longitudinal expansion rate, breaking the translational symmetry in the transverse directions can  change
the 1+1D picture only to the extent to which transverse gradients are not negligible compared to the longitudinal one. Therefore, at sufficiently early initialization, independent of
the transverse geometry and for all transverse positions $r$, arbitrary initial conditions in 3+1D evolve towards the 1+1D attractor.
% \textcolor{red}{The late-time evolution is insensitive to details of  the initial $v_z$-distribution (see \cite{Kurkela:2019kip}) and it thus does not depend on the initialization values of higher moments. This follows from the observation that irrespective of initial conditions, higher moments decay early to their attractor solution, see Ref.~\cite{Strickland:2019jut} and supplementary material.}
In contrast, the late-time evolution of the attractor does depend on the transverse profile of energy and transverse momentum.

\begin{figure}[h]
\begin{center}
 \includegraphics[width=0.45\textwidth]{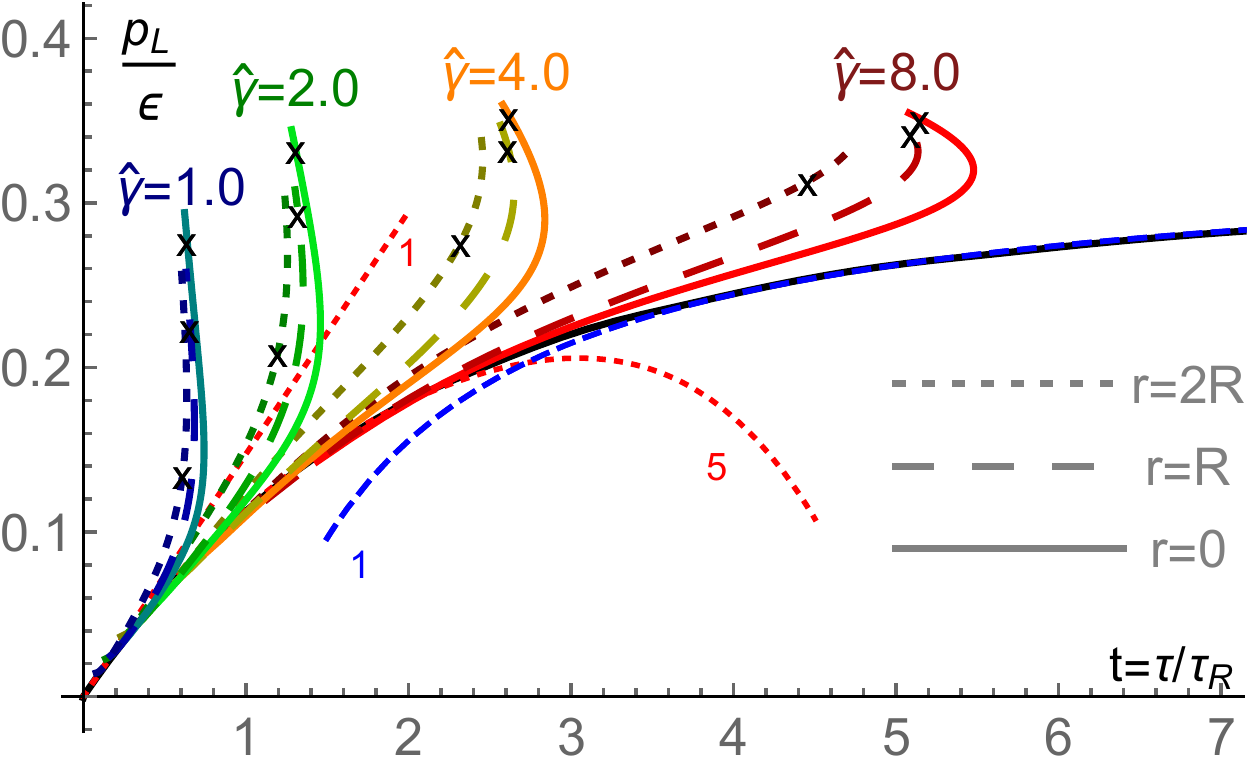}
\end{center}
\vspace{-.5cm}
  \caption{RTA attractor solutions of Bjorken-expanding 3+1D kinetic theory for collision systems of different opacity $\hat\gamma$ and three different transverse positions $r=0, R$, and $2R$.
  The 1+1D attractor (black line) corresponds to the limit of infinite opacity; the thin lines correspond to early- and late-time approximations, as in Fig.~\ref{fig2}. The black crosses denote the point on the
  attractor where the physical time reaches $\tau = 2R$ after which the system has decoupled \cite{Kurkela:2019kip}. 
  }
\label{fig3}
\end{figure}

These features are realized in boost-invariant 3+1D solutions of eq.~(\ref{eq10}), initialized with a Gaussian transverse energy profile with central energy density $\varepsilon_0$ and r.m.s. radius $R$, see Fig.~\ref{fig3}.
 For early initialization time $\tau_0$, keeping $\varepsilon_0\tau_0$ fixed, 
eq.~(\ref{eq10}) can be rescaled 
such that the evolution depends on only one dimensionless combination of model parameters, the opacity $\hat\gamma = \gamma R^{3/4} \left( \varepsilon_0\, \tau_0 \right)^{1/4} 
= \left(\gamma^3 \varepsilon_0^{3/4} R^3 t_0 \right)^{1/4}$, see~\cite{Kurkela:2018ygx,Kurkela:2019kip}. The opacity of a system increases with coupling strength ($\gamma$),
transverse system size ($R$) and initial central energy density ($\varepsilon_0$); physical collision systems were estimated to correspond to a range of opacities,  $\hat\gamma \lesssim 2$ for
proton-nucleus collision, $2 \lesssim \hat\gamma \lesssim 4$ for semi-peripheral PbPb collisions and somewhat higher values in central PbPb collisions~\cite{Kurkela:2019kip}.

The physical time in Fig.~\ref{fig3} is rescaled by a position- and time-dependent relaxation time $\tau_{\rm R}^{-1}(\tau,r) = \gamma \varepsilon(\tau,r)^{1/4}$. Therefore,
 for a system in which energy density decreases faster than $\propto \tau^{-4}$ due to transverse expansion, the relation between physical and rescaled time is not monotonic; this is the reason 
$t$ decreases for sufficiently late $\tau$ in the finite-$\hat\gamma$ curves of Fig.~\ref{fig3}. Moreover, because of this rescaling, 
 the deviation of the $r=0$ attractor solution from the 1+1D one, and the deviation of the attractor solutions at finite $r$ from the one at $r=0$ arise solely from the radial expansion. 
 For fixed $\hat\gamma$, the $r$-dependence is remarkably mild. Low orders in the early-time expansion are seen to be sufficient to describe systems characterized by 
values of $\hat\gamma$ that are within experimental reach. 
 What Fig.~\ref{fig3} makes abundantly clear is that what remains universal across collision geometries is not the late-time attractor but the early-time attractor. 
 That is, what remains universal is what follows from early-time dynamics and not what follows from hydrodynamization. 

In summary, we have studied the early-time behavior of qualitatively different models of collectivity applied to heavy-ion collisions. In some cases ($\mathcal{N}=4$ SYM), the attractor exists only at late times, and hydrodynamization and the loss of information about specific initial conditions are concurrent. The approach to the late-time attractor is then governed by the exponential decay of linearized non-hydrodynamic modes~\cite{Janik:2006gp,Spalinski:2018mqg}. 
In  other cases (IS and kinetic theory) a unique attractor extends to arbitrarily early times and thus specific information about the initial condition is lost well before hydrodynamization. 
We find that in this latter case, a qualitatively different power-law approach to the attractor is operational far-from-equilibrium. It is expansion-driven 
rather than interaction-driven. It is also noteworthy that in this respect, IS theory at any value of $\eta/s$ resembles RTA---the prototype of a weakly coupled system---rather than the prototypical
strongly coupled system of ${\cal N}=4$ SYM.

 % \emph{i.e.} the prototype of a weakly coupled system, 

One of the main challenges in heavy ion phenomenology is to elucidate the inner workings of the quark gluon plasma, and in particular, 
to discriminate between weakly and strongly coupled plasma models in which a quasi-particle picture exists or does not exist, respectively. 
The qualitative difference stated here between the early-time dynamics of strongly coupled ($\mathcal{N}=4$ SYM), compared to both,
kinetic theory and IS theory, deserves attention since it may help to make this distinction.

 % \newpage

\section{Supplemental Material}

\subsection{Series expansions of the IS attractor}

\begin{figure}[h]
\includegraphics[width=0.4\textwidth]{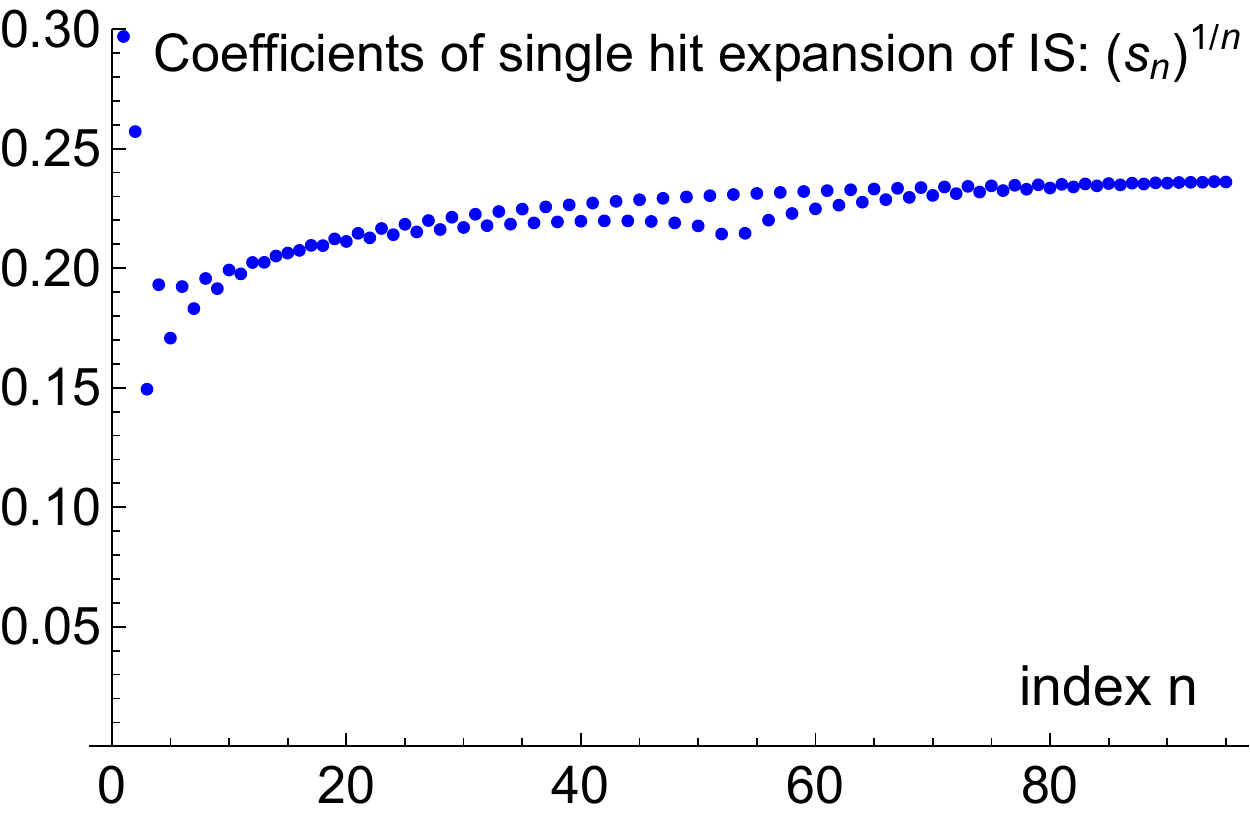}
\includegraphics[width=0.4\textwidth]{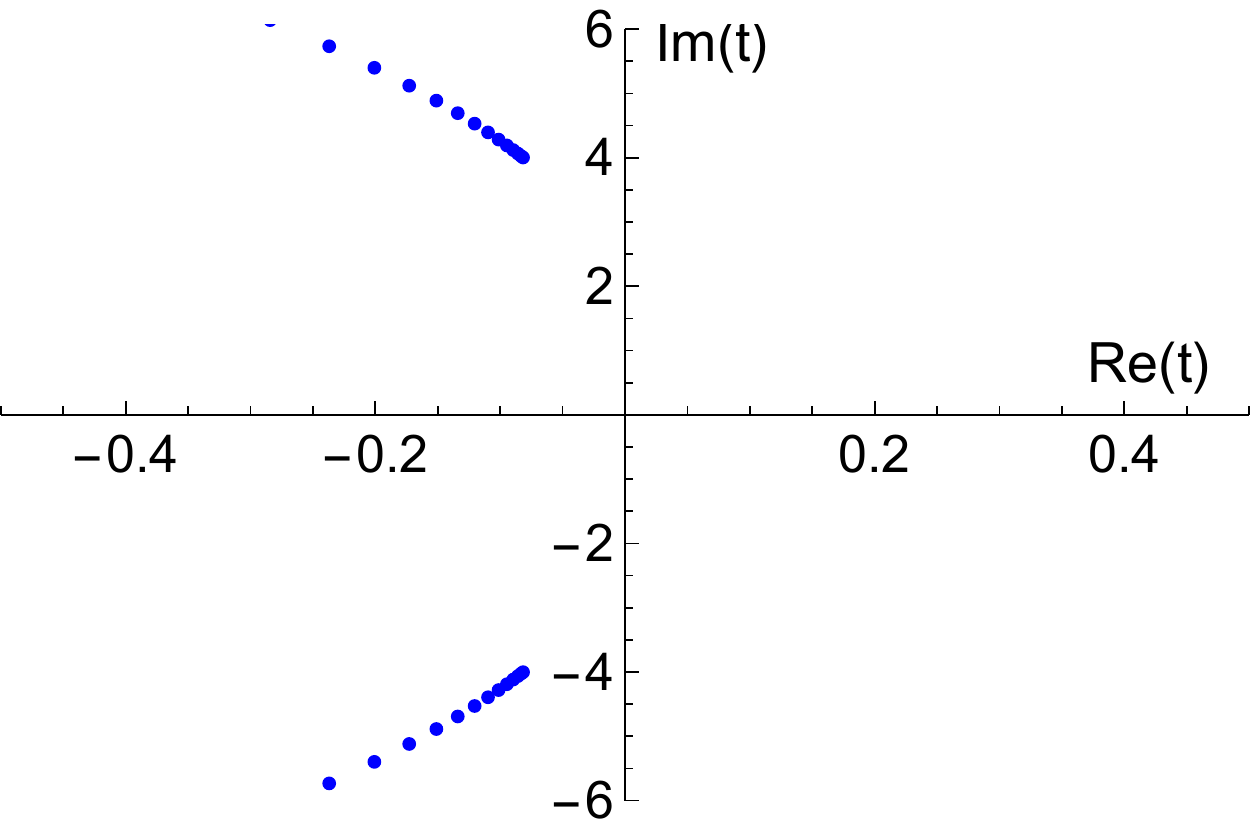}
\caption{Upper panel: The $n$'th root of the $n$'th order coefficient of the convergent single-hit expansion in the IS theory, eq.~(\ref{eq7}). Lower panel: The approximate analytic structure of the attractor solution. The dots correspond to the zeros of the denominator of the [50/50]-order Pad\'e-approximant of the series in eq.~(\ref{eq7}).}
\label{Pade}
\end{figure}

For Israel-Stewart theory, it is well-known that the late time (fluid-dynamic) expansion in powers of $1/t$,
\begin{align}
x_-(t) &= \frac{1}{3} + \sum_{i=1}^{\infty} \frac{h_i}{t^i}\, , \label{eq5}\\
h_1 = -\frac{1}{3} (16 a)&\, , h_2 = -\frac{176 a}{27}\, , \dots
% \\h_3&= \frac{32 a (144 a -121)}{243}
\end{align}
is an asymptotic, non-convergent but Borel-resummable series~[6]. 
% The asymptotic nature of this series is seen in Fig.~\ref{fig2}, where inclusion of successive orders worsens eventually 
% the agreement with the exact numerical solution at any finite $t$. Here, w
We contrast this hydrodynamic expansion with the corresponding early-time 
expansion of $x_-(t)$. Expansion in  powers of $t$ is possible
since $x_-(t)$ is regular for $t\to0$~[3, 17]
\begin{align}
x_-(t) &= x_-(0) + \sum_{i=1}^{\infty} s_i t^i\, ,  \label{eq7} \\
s_1 = \frac{4  \sqrt{5 a }}{9  \sqrt{ 5 a }+10} &, s_2 =  -\frac{60 \left(2 a+ \sqrt{ 5 a}\right)}{\left(\sqrt{5 a}+2\right) \left(9  \sqrt{ 5 a}+10\right)^2}\, ,  \dots
\nonumber
\end{align}
We have calculated these coefficients to high order. 
The coefficients of the series of eq.~(\ref{eq7}) are shown in the upper panel of Fig.~\ref{Pade}. The high-order coefficients determine the convergence properties of the series. That the high-order coefficients seem to saturate to $\sim 0.24^n$ suggests that the radius of convergence is that of a geometric series $\sum (0.24 t)^n$, that is $|t|\lesssim 4$. Consistent with that, the solution continued to complex $t$ shows non-analytic structures away from the real axis at $|t| \sim 4$.Therefore, unlike the late-time expansion, this early-time series is convergent, and inclusion of higher orders leads to a better agreement
within the radius of convergence $\approx 4$, see Fig.~2. Standard Pad\'e analysis shows that the convergence radius is set by a pair of poles at position $\approx \pm 4i$ in the complex $t$-plane, see Fig.~\ref{Pade}.
The early-time expansion can be analytically continued to any finite $t$ beyond its radius of convergence; the green-dashed line in Fig.~2 corresponds to a [10/10]-order Pad\'e  approximant 
that---in the displayed $t$-range---is indistinguishable within line-width from the full result. 

\subsection{Early-time power law decay to the RTA attractor}
For systems with transverse translational symmetry, the Boltzmann equation (7) can be reduced to a tower of moment equations 
(see e.g. [8, 18])
\begin{equation}
\partial_{\tau} p_{l}+\frac{1}{\tau}\left[(2 l+1) p_{l}-(2 l-1) p_{l+1}\right]=\frac{-1}{\tau_{\rm R}}\left[p_{l}-\frac{1}{2 l+1} \varepsilon\right]\, ,
\label{eq11}
\end{equation}
where $p_{l} \equiv \int_{-1}^{1} \frac{d v_{z}}{2} \int \frac{4 \pi d p p^{3}}{(2 \pi)^{3}} f  v_{z}^{2l} $; this definition implies $0 \leq \dots \leq p_2\leq p_1\leq p_0$.
According to (\ref{eq11}), evolution of $x=p_L/\varepsilon$ is coupled to higher moments. To characterize
the early-time power law decay to the attractor, we rewrite 
the hierarchy of RTA moment equations (\ref{eq11}) for $x_l\equiv p_l/\varepsilon$, 
\begin{eqnarray}
	\left(\frac{3}{4}-\frac{x}{4}\right) \frac{d x_l}{d t}&=&
%	\frac{4 \left(-3 x^2+6 x-3 y+t(3x-1)\right)}{3 t (x-3)}\, ,	
	- \frac{(2l-x)x_l -(2l-1) x_{l+1}}{t}	
\nonumber \\
&&	+\left(\frac{1}{(2l+1) }	- x_l\right)\, .
\label{eqa1}
\end{eqnarray}
Here, $x=x_1$; eq.~(8) is the first ($l=1$) of these moment equations.

\begin{figure}[h]
\begin{center}
  \includegraphics[width=0.4\textwidth]{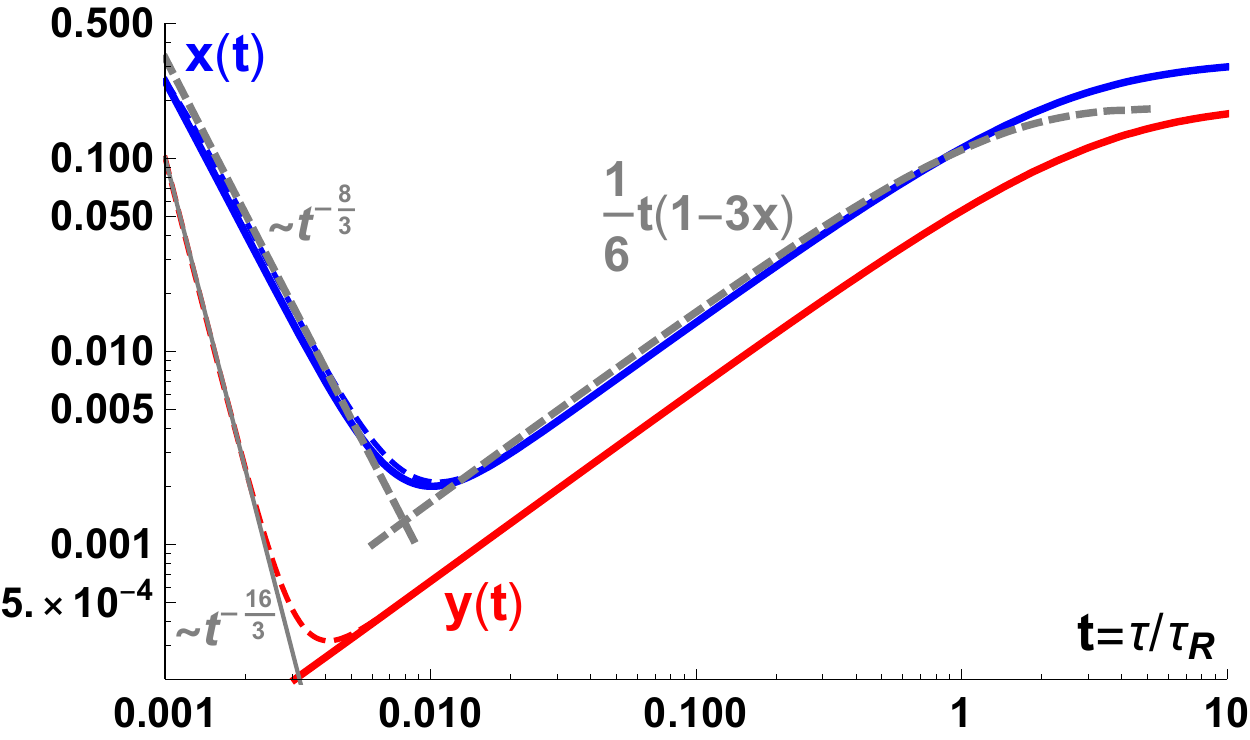}
\end{center}
  \caption{The early-time decay of $x(t)$ to the RTA attractor (thick blue line same as in Fig.~1) follows a power-law
  almost independent of the initial condition for the higher moment $y(t)$ (dashed and straight red line). This power law decay 
  seizes when the interaction rate becomes comparable to the expansion rate (see grey line and discussion in text).}
\label{figRTAapp}
\end{figure}

Fig.~\ref{figRTAapp} shows that the early-time power law decay of $x(t)$ towards the attractor becomes $\delta \sim t^{-8/3}$.
This late-time evolution is insensitive to details of the initial $v_z$-distribution (see [32]) and it thus does not depend on the initialization values of higher moments. We understand this numerical finding from the observation that irrespective of initial conditions, higher moments decay early to their attractor solution. For instance, the early-time decay of $y(t)$ is consistent with $\sim t^{-16/3}$, see also Fig.~\ref{figRTAapp}

According to the right hand side of (8), we expect the interaction rate (terms independent of $\propto t$) to balance the 
expansion rate ($\propto t^{-1}$) as soon as $x(t) \approx \textstyle\frac{1}{6} t \left(1- 3 x(t) \right)$. As seen from Fig.~\ref{figRTAapp},
it is at this time that the power law decay to the attractor seizes to persist. This further illustrates that expansion forces the decay to
the attractor prior to any contribution from the interaction rate.

In complete analogy to IS theory, also the late-time expansion of RTA kinetic theory is a non-convergent 
asymptotic and Borel-resummable series.  The early-time expansion results is a convergent series 
\begin{align}
x(t) = &\left(1-\frac{3 h_{-5}}{8}  \right) t \\
         &+ \frac{(15 h_{-5}-8) (h_{-5}-2 h_{-2})}{64} t^2 + \mathcal{O}(t^3)\nonumber, \\
         h_n = & \frac{4 \, _3F_2\left(\frac{1}{2},\frac{1}{2},1;\frac{3}{2},\frac{n}{8}+2;1\right)}{n+8}+\frac{4}{n+12}\, ,
\end{align}
that we have evaluated to high order. In complete analogy to IS theory, it can be analytically continued to any $t$ beyond its radius of convergence. Results of this expansion
are displayed in Fig.~2.

\subsection{A rapidly converging approximation to the IS attractor solution}
In the course of the present study, we stumbled upon a rapidly converging, analytic approximation.  While this solution is not needed for any step of the present paper, we document it in this
supplemental material.

\begin{figure}[b]
\begin{center}
    \includegraphics[width=0.4\textwidth]{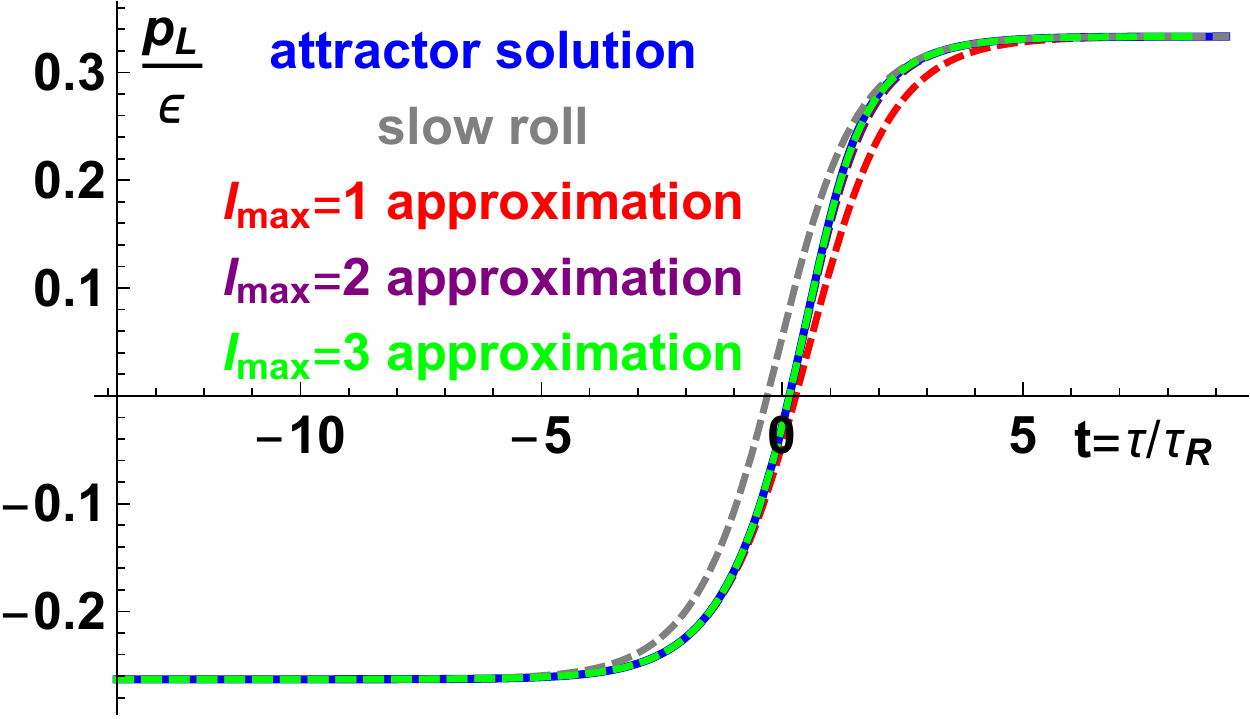}
  \includegraphics[width=0.4\textwidth]{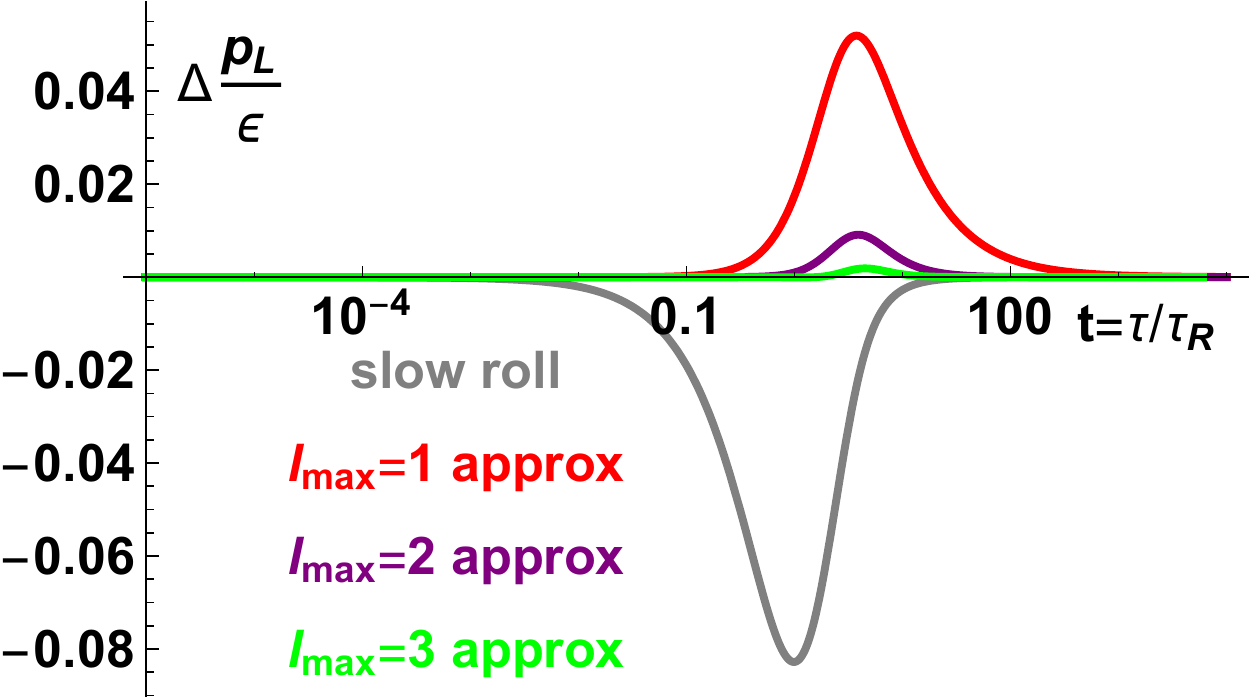}
\end{center}
  \caption{
 Upper panel: The IS attractor solution, compared to its slow roll expansion, and compared to the approximation $x(t) = c_0(t)$ where $c_0$ is determined from
  expanding (\ref{eqc1}) to first, second and third order respectively. Lower panel: the difference between the full IS attractor solution and the result obtained in
  the various approximations.}
\label{fighyd}
\end{figure}

The approximation is obtained by inserting into (\ref{eq3}) the Taylor series
\begin{equation}
	x_-(t) = \sum_{l=0}^{l_{\rm max}} c_l(t_*) \left( t-t_* \right)^l\, .
	\label{eqc1}
\end{equation}
 Here, expansion is around an arbitrary time $t_*$. Collecting powers of $t$ in (\ref{eq3}), one can express all coefficients $c_l$, $l\geq 1$, as rational
functions of $c_0$ and $t_*$. Since the derivative $dx/dt $ of the attractor solution must not diverge in the limit 
$t \to 0$, one requires
\begin{equation}
	x_-(t=0) = \frac{1}{15} \left(5 - 4 \sqrt{5} \right) = \sum_{l=0}^{l_{\rm max}} c_l(t_*) \left(-t_* \right)^l \, ,
	\label{eqc2}
\end{equation}
where the right-hand side is now an explicitly known rational function of $c_0(t_*)$ and $t_*$. Thus, 
(\ref{eqc2}) defines $c_0(t_*)$ implicitly in terms of $t_*$. By construction, see eq. (\ref{eqc1}),  
$c_0(t) = x(t)$ is the attractor solution, and (\ref{eqc2}) therefore provides an implicit expression for this solution.  
As seen in Fig.~\ref{fighyd}, even for a truncation at the lowest order $l_{\rm max} =1$, the accuracy of this procedure
is comparable to that of the slow-roll approximation \cite{Heller:2015dha}, and it improves rapidly upon including higher orders in the
truncated ansatz (\ref{eqc1}).

\end{document}